\documentclass[a4paper,11pt]{article}
\usepackage{pos}
\usepackage{gensymb}

\title{Particle Acceleration in the Cygnus Superbubble}
 \ShortTitle{Particle Acceleration in the Cygnus Superbubble}

\author*[a]{Binita Hona}
\forColl{HAWC}

\affiliation[a]{University of Utah \\
  Salt Lake City, UT, USA}

\emailAdd{binita.hona@utah.edu}

\abstract{The Cygnus Cocoon is the first gamma-ray superbubble powered by a massive stellar association, the OB2 association. It was postulated that the combined effects of the stellar winds of all the massive O-type stars of the OB2 association can accelerate the cosmic rays to PeV energy in the Cocoon. The conclusive proof of acceleration to PeV energy in the Cocoon will identify the stellar association as a PeV cosmic-ray accelerator, known as PeVatron. However, the Cocoon has been previously studied only up to 10 TeV. In this contribution, using 1343 days of High Altitude Water Cherenkov (HAWC) observatory data, we present the morphological and spectral study of the Cocoon above 1 TeV to beyond 100 TeV. The analysis at higher TeV energies reveals a softer spectrum compared to the GeV gamma-ray observation. This result suggests that the accelerator’s efficiency decreases around hundreds of TeV, or after being accelerated, the highest-energy protons escape the region. The study above 10 TeV presented here demonstrates how CR accelerators operate in these extreme energies and how particle transport impacts high-energy emission.}

\FullConference{37$^{\rm{th}}$ International Cosmic Ray Conference (ICRC 2021)\\
		July 12th -- 23rd, 2021\\
		Online -- Berlin, Germany}

\begin{document}
\maketitle

\section*{Introduction}

Stellar superbubbles, low gas density cavities formed by the interaction between the supersonic winds of massive type O stars are Galactic cosmic ray (CR) factories and are capable of accelerating cosmic rays to high energies \cite{1983SSRv...36..173C}. They might be the source of the CRs accelerated up to the "knee" of the CR spectrum or PeV acceleration in our Galaxy \cite{aha_nature}. The Cygnus region in our Galaxy contains massive molecular gas clouds, luminous HII regions and one of the closest giant star-forming complexes, called OB2 association. Hence, it is among the best places where gamma rays would be produced in association with the stellar clusters. The first stellar superbubble was discovered by Fermi-Large Area Telescope (LAT) at GeV energies in the Cygnus region \cite{fermicocoon}. The Fermi-LAT detected an extended region of gamma-ray emission between the two astrophysical objects, the OB2 association and the Gamma Cygni Supernova remnant (SNR). The Fermi-LAT study concluded that this extended emission is due to a single gamma-ray source rather than the multiple components \cite{fermicocoon}. The infrared emission surrounds the extended gamma-ray emission as in a stellar cocoon, hence the superbubble is also known as Fermi-LAT Cocoon. The morphology of the Cocoon emission is described by a symmetric Gaussian with a width of 2\degree  ~and the spectral energy distribution is described by a power law spectrum with a hard index of -2 \cite{fermicocoon}. The hard index implies that the CRs have not traveled too far from their source. Since there is no evidence of a shockwave from gamma Cygni SNR toward the direction of the superbubble, the SNR is unlikely to be the source powering the superbubble \cite{fermicocoon}. However, the collective effects of stellar winds from the massive stars in the OB2 association can explain presence of accelerated CRs in the Cocoon \cite{fermicocoon}. 

This GeV emission is co-located with a large extended region of emission detected by the Milagro observatory at TeV energy, MGRO J2031+41 \cite{milagro2007}. The previously unidentified MGRO J2031+4157 could be a TeV counterpart of the Fermi-LAT Cocoon \cite{fermicocoon}. The emission at TeV energies has also been confirmed by the ARGO-YBJ observatory. The ARGO J2031+41, a possible counterpart of the Fermi-LAT cocoon, is co-located with the GeV Cocoon and is described by a 2D Gaussian with a Gaussian width of 1.8\degree  ~\cite{argo}. The ARGO-YBJ collaboration reported the spectral measurements for this TeV Cocoon up to 10 TeV \cite{argo}. If we want to understand whether the stellar superbubble is capable of CR acceleration up to PeV energies, we need to study the gamma-ray emission beyond 10 TeV. In this study, using the data collected by the High Altitude Water Cherenkov (HAWC) observatory, we report the detection and spectral measurements of the TeV superbubble beyond 10 TeV \cite{hawc_cocoon}. 

The wide field of view HAWC TeV gamma-ray observatory is located at Sierra Negra, Puebla, Mexico at 4100 m. The main array of HAWC consists of 300 water Cherenkov detectors, each with 4 PMTs, and is sensitive to 300 GeV to beyond 100 TeV gamma rays. In the second HAWC catalog, 2HWC J2031+415 was detected near the Fermi-LAT Cocoon \cite{2hwccatalog}. For the study presented here, using 1343 days of HAWC data, a detailed analysis of the 2HWC J2031+415 region was performed to understand the TeV Cocoon component above 1 TeV. 

\section*{Analysis and Description of the Region}

The modelling of the region was done with the help of the publicly available multi-mission maximum likelihood (3ML) and HAWC accelerated Likelihood (HAL) software. The region of the interest (ROI) used for the study is a 6\degree  ~disk centered at the Fermi-LAT Cocoon location (RA = 307.17, Dec = 41.17)\degree ~ \cite{hawc_cocoon}. The nearby bright source 3HWC J2019+367 was excluded from this ROI by using a 2\degree ~disk mask. The single-source and multi-source models were tested for the ROI. The test statistics (TS) of a model fit is given by likelihood ratio test \cite{3ml}, $TS = 2 \ln \frac{L_{(\rm source)}}{L_{(\rm no-source)}}$, where~ $L_{(\rm source)}$ is the maximum likelihood of a model including a source and $L_{(\rm no-source)}$ is the maximum  likelihood of a model without the source in question. According to Wilks' theorem, the TS follows a $\chi^2$ distribution with the degree of freedom equal to the number of the free parameters in case of the nested models \cite{wilks}. The model with the best TS was chosen to describe the ROI. This model includes three sources as listed in Table \ref{tab:sources}. Based on the significance distribution of the residual map, this three source model is the best representation of the TeV gamma-ray emission in the ROI \cite{hawc_cocoon}. 

\begin{table*}[ht]
\centering

\begin{tabular}{c c c c} 
    \hline
    \textbf{Source} & \textbf{(RA, Dec)\degree} & \textbf{TS} & \textbf{Morphology}  \\ 
    
    \hline 
     &&&\\ 
      HAWC J2030+409 & (307.65, 40.93) & 195.2 & Gaussian width of 2.1\degree \\

      HAWC J2031+415 & (307.90, 41.51) & 298.5 & Gaussian width of 0.27\degree \\
      
      2HWC J2020+403 & (305.27, 40.52) & 53.7 & Disk radius of 0.63\degree \\
      &&&\\
    \hline  
\end{tabular}

\caption{\textbf{ The three sources in the ROI.}}
\label{tab:sources}
\end{table*}

In the model used to describe the ROI, 2HWC J2031+415 emission is disentangled into two sources: HAWC J2031+415 and HAWC J2030+409. HAWC J2031+415, a small extended Gaussian emission described by the Gaussian width of 0.27\degree, is possibly associated with a known pulsar wind nebula (PWN), TeV J2032+4130, which was first detected by HEGRA \cite{pwn2002} and has been later reported by other observatories such as MAGIC  \cite{MAGIC1}, WHIPPLE \cite{whipple1}, and VERITAS \cite{veritas2014}. HAWC J2030+409, a large extended source, is the TeV counterpart of the Cocoon. In addition to these two sources, there is also a third source in the ROI, 2HWC J2020+403, which lies 2.36\degree ~away from the centre of 2HWC J2031+415 location and is associated with the Gamma Cygni SNR \cite{verj20191}. The residual map after subtracting HAWC J2031+415 and 2HWC J2020+403 is shown in Figure \ref{fig:plot}.

\section*{Results: Morphology, Spectrum and Particle Modelling}

\begin{figure}
    \includegraphics[width=0.99\textwidth]{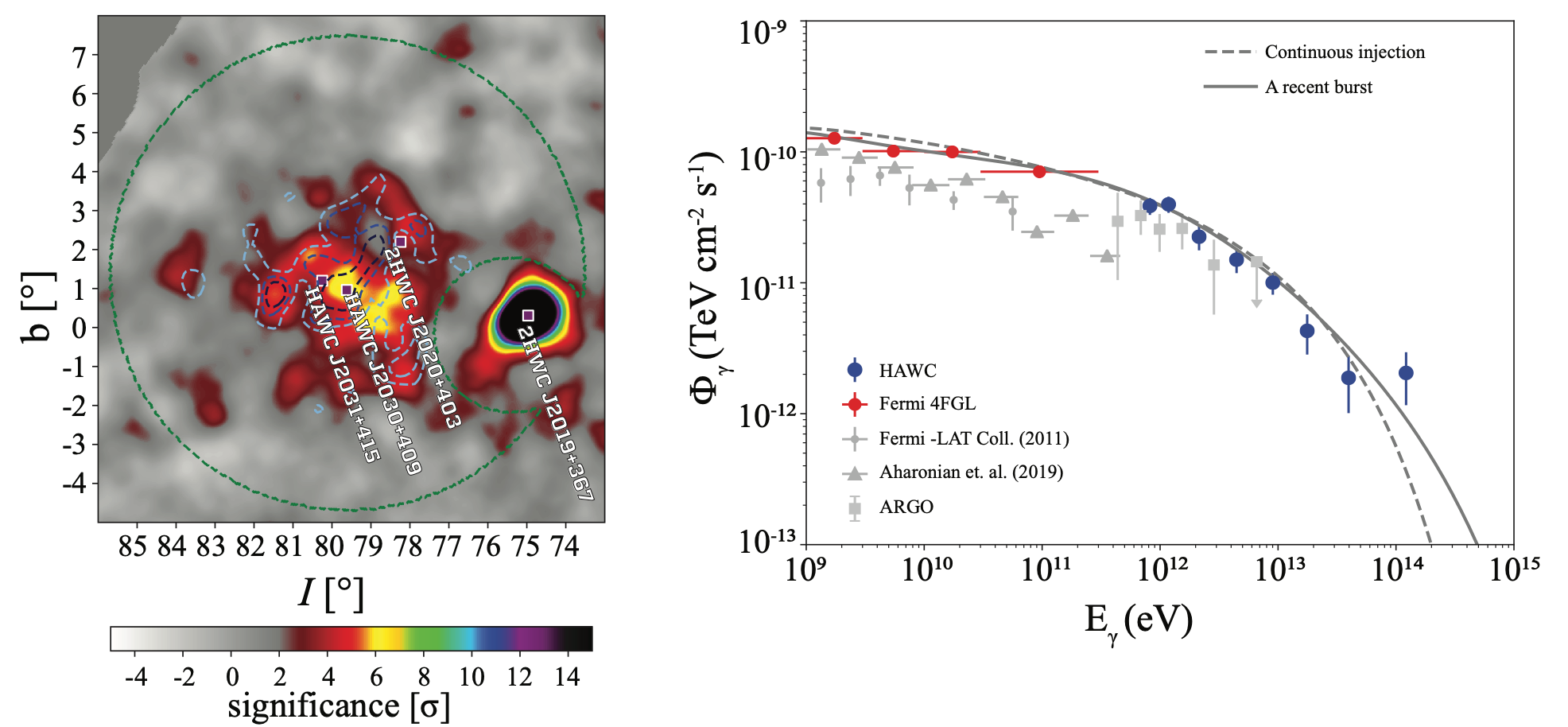}
    \caption{\textbf{Significance map and the spectral energy distribution of the gamma-ray emission of the HAWC Cocoon \cite{hawc_cocoon}.} \textbf{Left}: Significance map of the residual emission in the ROI after subtracting HAWC J2031+415 and 2HWC J2020+403.
 \textbf{Right}: Spectral energy distribution of the Cocoon measured by different $\gamma$-ray instruments, where the blue circles  are the spectral measurements for the Cocoon in this study. The errors on the flux points are the 1$\sigma$ statistical errors. The grey solid and dashed lines are $\gamma$-ray spectra derived from the hadronic modeling of the region.}
 \label{fig:plot}
\end{figure}

 HAWC J2030+409 or the HAWC Cocoon co-located with the Fermi-LAT Cocoon is detected with a TS of 195.2 at (RA = 307.65\degree \ $\pm$ 0.30\degree, Dec = 40.93\degree \ $\pm$ 0.26\degree). The emission morphology in HAWC data is described by a Gaussian profile with a width of 2.13\degree  $\pm$  0.15\degree . 
 The spectral energy distribution is described by a power-law spectrum where the pivot energy is 4.2 TeV. The flux normalization is $N_0=9.3_{-0.8}^{+0.9} (\rm stat.)  \times 10^{-13}  \rm \, \,cm^{-2}\,s^{-1}\,TeV^{-1}$ and the spectral index is $\Gamma = -2.64_{-0.05}^{+0.05} (\rm stat.)$ \cite{hawc_cocoon}. The energy range for the spectral measurement is 0.75 TeV to 210 TeV. The fluxes in 8 energy bins are shown in Figure \ref{fig:plot}. The flux obtained is compatible with the ARGO YBJ measurement at lower TeV energies \cite{argo} and with an extrapolation from the GeV measurement \cite{fermicocoon, 4fgl}. However, while the GeV spectrum has a hard spectral index of $\Gamma = -2.1$, a significant softening of the energy spectral density is observed at TeV range with $\Gamma = -2.6$ in both ARGO-YBJ up to 10 TeV \cite{argo}, and beyond 100~TeV with HAWC data \cite{hawc_cocoon}. 

A leptonic model assumed in \cite{hawc_cocoon} could not explain the GeV to TeV gamma-ray emission, however, two hadronic origin models were provided to explain the gamma-ray spectrum in the Cocoon and the spectral softening of gamma-ray emission at TeV energies \cite{hawc_cocoon}. The grey solid line is the expected gamma-ray distribution assuming a burst model, in which the acceleration process is powered by the recent starburst activities. Assuming a starburst activity less than 0.1~Myr ago and  a diffusion length in the Cocoon  100-1000 times shorter than that in the interstellar medium, the CRs are accelerated to PeV energies in the Cocoon \cite{hawc_cocoon}. The lower energy CRs remain in the cavity, whereas the highest energy CRs escape the region resulting in a spectral break from the GeV to the TeV regime \cite{hawc_cocoon}. The grey dotted line is the expected gamma-ray distribution assuming a continuous injection hadronic model, in which the acceleration process is powered by the stellar winds of multiple type O stars and has continuously injected particles over the OB2 lifetime (1 to 7 Myrs). The spectral softening at TeV energies is due to a cutoff in the injected CR spectrum. Assuming a diffusion length in the Cocoon 100-1000 times shorter than that in the interstellar medium, the cutoff in the injected CR spectrum is obtained at about 300 TeV \cite{hawc_cocoon}.

\section*{Discussion}
The  GeV observation by the Fermi-LAT observatory suggests that the OB2 association accelerates CRs and produces gamma-rays with a hard spectrum resulting from an efficient particle acceleration process such as the diffusive shock acceleration. The radial profile of the GeV gamma-ray emission implies the confinement of freshly accelerated CRs to the center of the Cocoon \cite{hawc_cocoon}. 

The analysis of HAWC data at higher TeV energies reveals new information about the Cocoon. In the HAWC band, the gamma-ray spectrum is described by a non-broken power law extending beyond 100 TeV which provides evidence that PeV cosmic rays were produced in this source. The flux measured by HAWC cannot be explained by inverse Compton scattering of electrons based on the leptonic modelling from  \cite{hawc_cocoon}. Additionally, the TeV Cocoon emission region overlaps with the GeV emission region. Considering the age of the Cocoon, this implies that the diffusion length in the Cocoon is much shorter than that in the interstellar medium. In comparison to GeV observation, the TeV gamma-ray spectrum softens significantly. This suggests that either the accelerator slows down around hundreds of TeVs (continuous injection hadronic model), or there is a leakage of the highest-energy protons from the region (burst hadronic model). In this study, the preference of one hadronic model over the other was not conclusive but might be obtained in the future with more statistics at even higher energy. 

The observation of the Cocoon above 10 TeV shows how the stellar superbubble operates at extreme energies and provides direct evidence that the superbubble can indeed produce CRs close to the knee region. The detailed study of the superbubble in the future data from the HAWC outriggers array, LHAASO observatory \cite{lhaaso}, and SWGO observatory \cite{swgo} will provide a clearer picture of the region.

\acknowledgments{We acknowledge the support from: the US National Science Foundation (NSF); the US Department of Energy Office of High-Energy Physics; the Laboratory Directed Research and Development (LDRD) program of Los Alamos National Laboratory; Consejo Nacional de Ciencia y Tecnolog\'ia (CONACyT), M\'exico, grants 271051, 232656, 260378, 179588, 254964, 258865, 243290, 132197, A1-S-46288, A1-S-22784, c\'atedras 873, 1563, 341, 323, Red HAWC, M\'exico; DGAPA-UNAM grants IG101320, IN111716-3, IN111419, IA102019, IN110621, IN110521; VIEP-BUAP; PIFI 2012, 2013, PROFOCIE 2014, 2015; the University of Wisconsin Alumni Research Foundation; the Institute of Geophysics, Planetary Physics, and Signatures at Los Alamos National Laboratory; Polish Science Centre grant, DEC-2017/27/B/ST9/02272; Coordinaci\'on de la Investigaci\'on Cient\'ifica de la Universidad Michoacana; Royal Society - Newton Advanced Fellowship 180385; Generalitat Valenciana, grant CIDEGENT/2018/034; Chulalongkorn University’s CUniverse (CUAASC) grant; Coordinaci\'on General Acad\'emica e Innovaci\'on (CGAI-UdeG), PRODEP-SEP UDG-CA-499; Institute of Cosmic Ray Research (ICRR), University of Tokyo, H.F. acknowledges support by NASA under award number 80GSFC21M0002. We also acknowledge the significant contributions over many years of Stefan Westerhoff, Gaurang Yodh and Arnulfo Zepeda Dominguez, all deceased members of the HAWC collaboration. Thanks to Scott Delay, Luciano D\'iaz and Eduardo Murrieta for technical support.}

\clearpage
\bibliographystyle{plain}
\bibliography{bibliography}


\clearpage
\section*{Full Authors List: \Coll\ Collaboration}


\scriptsize
\noindent
A.U. Abeysekara$^{48}$,
A. Albert$^{21}$,
R. Alfaro$^{14}$,
C. Alvarez$^{41}$,
J.D. Álvarez$^{40}$,
J.R. Angeles Camacho$^{14}$,
J.C. Arteaga-Velázquez$^{40}$,
K. P. Arunbabu$^{17}$,
D. Avila Rojas$^{14}$,
H.A. Ayala Solares$^{28}$
% ...

\noindent
% ...
$^{14}$Instituto de Física, Universidad Nacional Autónoma de México, Ciudad de México, México,
% ...
$^{17}$Instituto de Geofísica, Universidad Nacional Autónoma de México, Ciudad de México, México,
% ...
$^{21}$Physics Division, Los Alamos National Laboratory, Los Alamos, NM, USA,
% ...
$^{28}$Department of Physics, Pennsylvania State University, University Park, PA, USA,
% ...
$^{40}$Instituto de Física y Matemáticas, Universidad Michoacana de San Nicolás de Hidalgo, Morelia, Michoacán, México,
$^{41}$FCFM-MCTP, Universidad Autónoma de Chiapas, Tuxtla Gutiérrez, Chiapas, México,
% ...
$^{48}$Department of Physics and Astronomy, University of Utah, Salt Lake City, UT, USA,
% ...

%
%
%

\end{document}